# Exact solutions of a class of double-well potentials: Algebraic Bethe ansatz

*M. Baradaran[1] and H. Panahi[2]*

*Department of Physics, University of Guilan, Rasht 41635-1914, Iran*

**Abstract**

In this paper, applying the Bethe ansatz method, we investigate the Schrödinger equation for the three quasi-exactly solvable double-well potentials, namely the generalized Manning potential, the Razavy bistable potential and the hyperbolic Shifman potential. General exact expressions for the energies and the associated wave functions are obtained in terms of the roots of a set of algebraic equations. Also, we solve the same problems using the Lie algebraic approach of quasi-exact solvability through the $sl(2)$ algebraization and show that the results are the same. The numerical evaluation of the energy spectrum is reported to display explicitly the energy levels splitting.

## 1. Introduction

Double well potentials (DWPs) are an important class of configurations have been extensively used in many fields of physics and chemistry for the description of the motion of a particle under two centers of force. Recently, solutions of the Schrödinger equation with DWPs have found applications in the Bose–Einstein condensation [1], molecular systems [2], quantum tunneling effect [3,4], microscopic description of Tunneling Systems [5] and etc. Some well-known DWPs in the literature are the quartic potential [6], the sextic potential [7], the Manning potential [2] and the Razavy potential [8]. In addition, it has been found that with some special constraints on the parameters of these potentials, a finite part of the energy spectrum and corresponding eigenfunctions can be obtained as explicit expressions in a closed form. In other words, these systems are quasi-exactly solvable (QES) [9–13]. DWPs in the framework of QES systems have received a great deal of attention. This is due to the pioneering work of Razavy, who proposed his well-known potential for describing the quantum theory of molecules [8]. The fundamental idea behind the quasi-exact solvability is the existence of a hidden dynamical symmetry. QES systems can be studied by two main approaches: the analytical approach based on the Bethe ansatz [14-19] and the Lie algebraic approach [10-13]. These techniques, are of great importance because only a few number of problems in quantum mechanics can be solved exactly. Therefore, these approaches can be applied as accurate and efficient techniques to study and solve the new problems that arise in different areas of physics such as quantum field theory [20-22], condensed matter physics [23-25], quantum cosmology [26-32] and so on, which their exact solutions are hard to obtain or are impossible to find. In the literature, DWPs have been studied by using various techniques such as the WKB approximation [33,34], asymptotic iteration method (AIM) [35], the Wronskian method [36], etc. On the other hand, it is well-known that the tunnel splitting which is the differences between the adjacent energy levels, is the characteristic of the energy spectrum for the DWPs [37-40]. In this paper, we apply two different methods to solve the Schrödinger equation for three QES DWPs: the Bethe ansatz method (BAM) and the Lie algebraic method, and show that the results of the two methods are consistent. Also, we provide some numerical results of the bistable Razavy potential to display the energy levels splitting explicitly.

[1] Corresponding author E-mail: marzie.baradaran@yahoo.com

[2] E-mail: t-panahi@guilan.ac.ir

This paper is organized as follows: In section 2, we introduce the QES DWPs and obtain the exact solutions of the corresponding Schrödinger equations using the BAM. Also, general exact expressions for the energies and the wave functions are obtained in terms of the roots of the Bethe ansatz equations. In section 3, we solve the same problems using the Lie algebraic approach within the framework of quasi-exact solvability and therein we make a comparison between the solutions obtained by the BAM and QES method. We end with conclusions in section 4. In the Appendix, we review the connection between *sl*(2) Lie algebra and the second order QES differential equations.

## 2. The BAM for the DWPs

In this section, we introduce the three DWPs that are discussed in this work, and solve the corresponding Schrödinger equations via the factorization method in the framework of algebraic Bethe ansatz [15]. The general exact expressions for the energies, the wave functions and the allowed values of the potential parameters are obtained in terms of the roots of the Bethe ansatz equations.

### 2.1. The generalized Manning potential

First, we consider the three parameter generalized Manning potential as [36]

$$V(x) = -v_1 \sech^6(x) - v_2 \sech^4(x) - v_3 \sech^2(x). \tag{1}$$

The parameters $v_1$, $v_2$ and $v_3$ are real constants which under certain constraint conditions enable us to obtain the bound-state eigenenergies and associated wave functions exactly. In atomic units ($m = \hbar = c = 1$), the Schrödinger equation with potential (1) is

$$\left(-\frac{d^2}{dx^2}-v_1\sech^6(x)-v_2\sech^4(x)-v_3\sech^2(x)\right)\psi(x)=E\psi(x). \qquad (2)$$

Xie [36] has studied this problem and obtained exact solutions of the first two states in terms of the confluent Heun functions. In this paper, we intend to extend the results of Ref. [36] by determining general exact expressions for the energies, wave functions and the allowed values of the potential parameters, using the factorization method in the framework of the Bethe ansatz. To this end, and for the purpose of extracting the asymptotic behaviour of the wave function, we consider the following transformations

$$z=\tanh^2(x),$$
$$\psi(x)=(1-z)^{\frac{\sqrt{-E}}{2}}e^{\frac{\sqrt{v_1}}{2}z}\phi(z), \qquad (3)$$

which after substituting into Eq. (2), gives

$$H\phi(z)=0,$$
$$H=-z(z-1)\frac{d^2}{dz^2}-\left(\sqrt{v_1}z^2+\left(\frac{3}{2}+\sqrt{-E}-\sqrt{v_1}\right)z-\frac{1}{2}\right)\frac{d}{dz}-\frac{1}{4}(\lambda z+\varepsilon), \qquad (4)$$

where

$$\lambda=(3+2\sqrt{-E})\sqrt{v_1}+v_1+v_2,$$
$$\varepsilon=-v_1-v_2-v_3-E+\sqrt{-E}-\sqrt{v_1}. \qquad (5)$$

In order to solve the present problem via BAM, we try to factorize the operator $H$ as

$$H=A_n^+A_n, \qquad (6)$$

such that $A_n\phi_n(z)=0$. Now, we suppose that polynomial solution (Bethe ansatz) exist for (4) as

$$\phi_n(z)=\begin{cases}\prod_{k=1}^{n}(z-z_k) & n\neq 0 \\ 1 & n=0\end{cases}, \qquad (7)$$

with the distinct roots $z_k$ that are interpreted as the wave function nodes and can be determined by the Bethe ansatz equations. As a result, it is evident that the operator $A_n$ must have the form

$$A_n=\frac{d}{dz}-\sum_{k=1}^{n}\frac{1}{z-z_k}, \qquad (8)$$

and then, the operator $A_n^+$ has the following form

$$A_n^+=-\frac{d}{dz}-\left(\frac{1}{2z}+\frac{1+\sqrt{-E}}{z-1}+\sqrt{v_1}\right)-\sum_{k=1}^{n}\frac{1}{z-z_k}. \qquad (9)$$

Substituting (8) and (9) into (6), we have

$$A_n^+ A_n = -\frac{d^2}{dz^2} - \left(\frac{2\sqrt{v_1}z^2 + (3 + 2\sqrt{-E} - 2\sqrt{v_1})z - 1}{2z(z-1)}\right)\frac{d}{dz}$$

$$+ \left(\frac{1}{2z} + \frac{1+\sqrt{-E}}{z-1} + \sqrt{v_1} + \sum_{j \neq k}^n \frac{2}{z_j - z_k}\right)\sum_{k=1}^n \frac{1}{z - z_k}. \tag{10}$$

The last term on the right of Eq. (10) is obviously a meromorphic function with simple poles at $z = 0, 1$ and $z_k$. Comparing the treatment of Eq. (10) with Eq. (4) at these points, we obtain the following relations for the unknown roots $z_k$ (the so-called Bethe ansatz equation), the energy eigenvalues and the constraints on the potential parameters

$$\sum_{j \neq k}^n \frac{2}{z_k - z_j} + \frac{1}{2z_k} + \frac{1 + \sqrt{-E_n}}{z_k - 1} + \sqrt{v_1} = 0, \tag{11}$$

$$(3 + 2\sqrt{-E_n})\sqrt{v_1} + v_1 + v_2 + 4n\sqrt{v_1} = 0, \tag{12}$$

$$-v_1 - v_2 - v_3 - E_n + \sqrt{-E_n} - \sqrt{v_1} + 4\sqrt{v_1}\sum_{k=1}^n z_k + 4n(n-1) + 4\left(\frac{3}{2} + \sqrt{-E_n} - \sqrt{v_1}\right)n = 0. \tag{13}$$

As examples of the above general solutions, we study the ground, first and second excited states of the model in detail. For $n = 0$, by Eqs. (12) and (3), we have the following relations

$$(3 + 2\sqrt{-E_0})\sqrt{v_1} + v_1 + v_2 = 0,$$

$$\psi_0(z) = (1-z)^{\frac{\sqrt{-E_0}}{2}} e^{\frac{\sqrt{v_1}}{2}z}, \tag{14}$$

for the ground state energy and wave function, with the potential constraint given by

$$-v_1 - v_2 - v_3 - E_0 + \sqrt{-E_0} - \sqrt{v_1} = 0. \tag{15}$$

For the first excited state $n = 1$, by Eqs. (12) and (3), we have

$$(3 + 2\sqrt{-E_1})\sqrt{v_1} + v_1 + v_2 + 4\sqrt{v_1} = 0,$$

$$\psi_1(z) = (1-z)^{\frac{\sqrt{-E_1}}{2}} e^{\frac{\sqrt{v_1}}{2}z} (z - z_1), \tag{16}$$

for the energy and wave function, respectively. Also, the constraint condition between the parameters of the potential is as

$$-v_1 - v_2 - v_3 - E_1 + 4\sqrt{v_1}z_1 + 5\sqrt{-E_1} - 5\sqrt{v_1} + 6 = 0, \tag{17}$$

where the root $z_1$ is obtained from the Bethe ansatz equation (11) as

$$z_1 = \frac{-\frac{3}{2} - \sqrt{-E_1} + \sqrt{v_1} \pm \sqrt{\left(\frac{3}{2} + \sqrt{-E_1} - \sqrt{v_1}\right)^2 + 2\sqrt{v_1}}}{2\sqrt{v_1}}. \tag{18}$$

Similarly, for the second excited state $n = 2$, the energy, wave function and the constraint condition between the potential parameters are given as

$$\left(3+2\sqrt{-E_2}\right)\sqrt{v_1}+v_1+v_2+8\sqrt{v_1}=0, \tag{19}$$

$$\psi_2(z)=(1-z)^{\frac{\sqrt{-E_2}}{2}}e^{\frac{\sqrt{v_1}}{2}z}(z^2-(z_1+z_2)z+z_1z_2), \tag{20}$$

$$-v_1-v_2-v_3-E_2+4\sqrt{v_1}(z_1+z_2)+9\sqrt{-E_2}-9\sqrt{v_1}+20=0, \tag{21}$$

where the two distinct roots $z_1$ and $z_2$ are obtainable from the Bethe ansatz equations

$$\begin{aligned}\frac{2}{z_1-z_2}+\frac{1}{2z_1}+\frac{1+\sqrt{-E_2}}{z_1-1}+\sqrt{v_1}=0,\\\frac{2}{z_2-z_1}+\frac{1}{2z_2}+\frac{1+\sqrt{-E_2}}{z_2-1}+\sqrt{v_1}=0.\end{aligned} \tag{22}$$

In Table 1, we report and compare our numerical results for the first three states. Also, in Fig. 1, we draw the potential (1) for the possible values of the parameters $v_1=1$, $v_2=-12$ and $v_3=15.255$.

### 2.2. The Razavy bistable potential

Here, we consider the hyperbolic Razavy potential (also called the double sinh-Gordon (DSHG) potential) defined by [41]

$$V(x)=\left(\xi\cosh(2x)-M\right)^2, \tag{23}$$

where $\xi$ is a real parameter. While the value of $M$ is not restricted generally, but according to Ref. [8], the solutions of the first $M$ states can be found exactly if $M$ is a positive integer. This potential exhibits a double-well structure for $M>\xi$ with the two minima lying at $\cosh(2x_0)=M/\xi$. Specifically, the Razavy potential can be considered as a realistic model for a proton in a hydrogen bond [42,43]. The potential (23) has also been used by several authors, for studying the statistical mechanics of DSHG kinks theory [44]. The Schrödinger equation with potential (23) is

$$\left(-\frac{d^2}{dx^2}+\left(\xi\cosh(2x)-M\right)^2\right)\psi(x)=E\psi(x). \tag{24}$$

Exact and approximate solutions of the first $M$ states for $M=1,2,3,4,5,6,7$ has obtained via different methods and can be found in Refs. [8,35,44]. In this and the next section, we extend the solutions of (24) to the general cases of arbitrarily $M$ and obtain general exact expressions for the energies and wave functions using the BAM and QES methods. Using the change of variable $z=e^{2x}$ and the gauge transformation

$$\psi(z)=z^{\frac{1-M}{2}}e^{-\frac{\xi}{4}\left(z+\frac{1}{z}\right)}\phi(z), \tag{25}$$

we obtain

$$H\phi(z)=0,$$
$$H=-4z^2\frac{d^2}{dz^2}+\left(2\xi z^2+(4M-8)z-2\xi\right)\frac{d}{dz}+\left(-2\xi(M-1)z+\xi^2+2M-1-E\right). \tag{26}$$

Now, we consider the polynomial solutions for (26) as

$$\phi_M(z) = \begin{cases} \prod_{k=2}^{M}(z-z_k) & M \neq 1 \\ 1 & M = 1 \end{cases}, \quad (27)$$

where $z_k$ are unknown parameters to be determined by the Bethe ansatz equations. In this case, the operators $A_M$ and $A_M^+$ are defined as

$$A_M = \frac{d}{dz} - \sum_{k=2}^{M}\frac{1}{z-z_k},$$

$$A_M^+ = -\frac{d}{dz} + \frac{\xi}{2} + \frac{M-2}{z} - \frac{\xi}{2z^2} - \sum_{k=2}^{M}\frac{1}{z-z_k}. \quad (28)$$

As a result, we have

$$A_M^+ A_M =$$

$$-\frac{d^2}{dz^2} + \left(\frac{2\xi z^2 + 4(M-2)z - 2\xi}{4z^2}\right)\frac{d}{dz} + \left(-\frac{\xi}{2} - \frac{M-2}{z} + \frac{\xi}{2z^2} + \sum_{j\neq k}^{M}\frac{2}{z_j - z_k}\right)\sum_{k=2}^{M}\frac{1}{z-z_k}. \quad (29)$$

Now, evaluating the residues at the two simple poles $z = z_k$ and $z = 0$, and comparing the results with (26), we obtain the following relations

$$E_M = \xi^2 + 2M - 1 + 2\xi \sum_{k=2}^{M} z_k, \quad (30)$$

$$\sum_{j\neq k}^{M}\frac{2}{z_k - z_j} + \frac{\xi}{2z_k^2} - \frac{M-2}{z_k} - \frac{\xi}{2} = 0, \quad (31)$$

for the energy eigenvalues and the roots $z_k$, respectively. For example, for the ground state $M = 1$, from Eqs. (30) and (25), we have the following relations for energy and wave function

$$E_1 = \xi^2 + 1, \quad (32)$$

$$\psi_1(z) = e^{-\frac{\xi}{4}\left(z+\frac{1}{z}\right)}. \quad (33)$$

For the first excited state $M = 2$, from Eqs. (30) and (25), we have

$$E_2 = \xi^2 + 3 + 2\xi(z_2), \quad (34)$$

$$\psi_2(z) = z^{\frac{-1}{2}} e^{-\frac{\xi}{4}\left(z+\frac{1}{z}\right)}(z-z_2), \quad (35)$$

where $z_2 = \pm 1$. Similarly, the second excited state solution corresponds to the $M = 3$ are given by

$$E_3 = \xi^2 + 5 + 2\xi(z_2 + z_3), \quad (36)$$

$$\psi_3(z) = z^{-1} e^{-\frac{\xi}{4}\left(z+\frac{1}{z}\right)}\left(z^2 - (z_2+z_3)z + z_2 z_3\right), \quad (37)$$

where the distinct roots $z_2$ and $z_3$ are obtained from the Bethe ansatz equation (31) as follows

$$\begin{pmatrix} z_2 \\ z_3 \end{pmatrix} = \begin{cases} \begin{pmatrix} \pm 1 \\ \mp 1 \end{pmatrix}, \\ \\ \begin{pmatrix} \dfrac{\pm\left(\sqrt{2}\sqrt{1+\sqrt{4\xi^2+1}} \pm \sqrt{4\xi^2+1} \pm 1\right)\sqrt{2}\sqrt{1+\sqrt{4\xi^2+1}}\left(-1+\sqrt{4\xi^2+1}\right)}{2\left(\sqrt{2}\left(\sqrt{4\xi^2+1}+3\right)\sqrt{1+\sqrt{4\xi^2+1}} \pm 4\sqrt{4\xi^2+1} \pm 4\right)\xi} \\ \dfrac{\pm\sqrt{2}\sqrt{1+\sqrt{4\xi^2+1}} + \sqrt{4\xi^2+1}+1}{2\xi} \end{pmatrix}, \\ \\ \begin{pmatrix} \dfrac{-\left(-1+\sqrt{4\xi^2+1} \mp \sqrt{2-2\sqrt{4\xi^2+1}}\right)\sqrt{2-2\sqrt{4\xi^2+1}}\left(1+\sqrt{4\xi^2+1}\right)}{2\xi\left(\left(\sqrt{4\xi^2+1}-3\right)\sqrt{2-2\sqrt{4\xi^2+1}} \pm 4\sqrt{4\xi^2+1} \mp 4\right)} \\ \dfrac{1-\sqrt{4\xi^2+1} \pm \sqrt{2-2\sqrt{4\xi^2+1}}}{2\xi} \end{pmatrix}. \end{cases} \quad (38)$$

The results obtained for the first three levels are reported and compared in table 2. The Razavy potential and its energy levels splitting are plotted in Fig. 2. Also, the numerical results for the eigenvalues and energy levels splitting are presented in table 3. As can be seen, for a given $M$, the energy differences between the two adjacent levels satisfy the inequality $E_2 - E_1 < E_4 - E_3 < \ldots$ and therefore the energy levels are paired together.

### 2.3. The hyperbolic Shifman potential

Now, we consider a hyperbolic potential introduced by Shifman as [9]

$$V(x) = \frac{a^2}{2}\sinh^2(x) - a\left(n+\frac{1}{2}\right)\cosh(x), \quad (39)$$

where the parameter $a$ is a real constant. The Schrödinger equation for potential (39) is given by

$$\left(-\frac{d^2}{dx^2} + \frac{a^2}{2}\sinh^2(x) - a\left(n+\frac{1}{2}\right)\cosh(x) - E\right)\psi(x) = 0. \quad (40)$$

According to the asymptotic behaviors of the wave function at the origin and infinity, we consider the following transformations

$$\begin{aligned}\psi(x) &= e^{-a\cosh(x)}\phi(x), \\ z &= \cosh(x).\end{aligned} \quad (41)$$

Therefore, the differential equation for $\varphi(x)$ reads

$$H\phi(z) = 0,$$
$$H = (-\frac{1}{2}z^2 + \frac{1}{2})\frac{d^2}{dz^2} + \left(az^2 - \frac{1}{2}z - a\right)\frac{d}{dz} - (naz + E). \tag{42}$$

Now, by assuming

$$\phi_n(z) = \begin{cases} \prod_{k=1}^{n}(z - z_k) & n \neq 0 \\ 1 & n = 0 \end{cases}, \tag{43}$$

and defining the operators $A_n$ and $A_n^+$ as

$$A_n = \frac{d}{dz} - \sum_{k=1}^{n}\frac{1}{z - z_k},$$
$$A_n^+ = -\frac{d}{dz} - \left(\frac{2az^2 - z - 2a}{1 - z^2}\right) - \sum_{k=1}^{n}\frac{1}{z - z_k}, \tag{44}$$

we obtain

$$A_n^+ A_n = -\frac{d^2}{dz^2} - \left(\frac{2az^2 - z - 2a}{1 - z^2}\right)\frac{d}{dz} + \left(\frac{2az^2 - z - 2a}{1 - z^2} + \sum_{j \neq k}\frac{2}{z_j - z_k}\right)\sum_{k=1}^{n}\frac{1}{z - z_k}. \tag{45}$$

Comparing the residues at the simple poles $z = \pm 1$ and $z = z_k$ with (42), we obtain the following set of equations for the energy and the zeros $z_k$

$$E_n = a\sum_{k=1}^{n}z_k - \frac{n^2}{2}, \tag{46}$$

$$\sum_{j \neq k}\frac{2}{z_j - z_k} + \frac{2az_k^2 - z_k - 2a}{1 - z_k^2} = 0, \tag{47}$$

respectively. Here, we obtain exact solutions of the first three levels. For $n = 0$, from Eqs. (46) and (41), we get

$$E_0 = 0, \tag{48}$$
$$\psi_0(z) = e^{-az}, \tag{49}$$

and for the first excited state $n = 1$,

$$E_1 = az_1 - \frac{1}{2}, \tag{50}$$
$$\psi_1(z) = e^{-az}(z - z_1), \tag{51}$$

where the root $z_1$ is obtained from Bethe ansatz equation (47) as

$$z_1 = \frac{1 \pm \sqrt{1 + 16a^2}}{4a}. \tag{52}$$

Solutions of the second excited state corresponds to $n = 2$ are given as

$$E_2 = a(z_1 + z_2) - 2, \tag{53}$$

$$\psi_2(z) = e^{-az}\left(z^2 - (z_1 + z_2)z + z_1 z_2\right), \tag{54}$$

where the roots $z_1$ and $z_2$ are obtained from Eq. (47) as

$$(z_1, z_2) = \begin{cases} (-0.7378701975, 0.6708350007), \\ (17.31284254, 3.232902445), \\ (-0.3019410394, 14.82323126). \end{cases} \tag{55}$$

Here, we have taken the parameter $a = 0.1$. Our numerical results obtained for the first three levels are displayed and compared in table 4. Also, the Shifman DWP for the parameter values $a = 0.1$ and $n = 1$ is plotted in Fig. 3. In the next section, we intend to reproduce the results using the Lie algebraic approach in the framework of quasi-exact solvability.

## 3. The Lie algebraic approach for the DWPs

In the previous section, we applied the BAM to obtain the exact solutions of the systems. In this section, we solve the same models by using the Lie algebraic approach and show how the relation with the $sl(2)$ Lie algebra underlies the solvability of them. To this aim, for each model, we show that the corresponding differential equation is an element of the universal enveloping algebra of $sl(2)$ and thereby we obtain the exact solutions of the systems using the representation theory of $sl(2)$. The method is outlined in the Appendix.

### 3.1. The generalized Manning potential

Applying the results of Appendix, it is easy to verify that Eq. (4) can be written in the Lie algebraic form

$$H\phi(z) = 0,$$
$$H = -J_n^+ J_n^- - J_n^0 J_n^- - \sqrt{v_1} J_n^+ + \left(\frac{3}{2} + n + \sqrt{-E} - \sqrt{v_1}\right) J_n^0 - \left(\frac{n+1}{2}\right) J_n^- + \left(\frac{n^2}{2} + \left(\frac{3}{2} + \sqrt{-E} - \sqrt{v_1}\right)\frac{n}{2} + \frac{\varepsilon}{4}\right), \tag{56}$$

if the following condition (constraint of quasi-exact solvability) is fulfilled

$$(3 + 2\sqrt{-E_n})\sqrt{v_1} + v_1 + v_2 = -4n\sqrt{v_1}, \tag{57}$$

which is the same result as Eq. (12). As a result, the operator $H$ preserves the finite-dimensional invariant subspace $\phi(z) = \sum_{m=0}^{n} a_m z^m$ spanned by the basis $\langle 1, z, z^2, ..., z^n \rangle$ and therefore the $n+1$ states can be determined exactly. Accordingly, Eq. (56) can be represented as a matrix equation whose non-trivial solution exists if the following constraint is satisfied (Cramer's rule)

$$\begin{vmatrix} \dfrac{\varepsilon}{4} & -\dfrac{1}{2} & 0 & 0 & & 0 \\ -n\sqrt{v_1} & \sqrt{-E}-\sqrt{v_1}+\dfrac{6+\varepsilon}{4} & -3 & 0 & & 0 \\ 0 & -(n-1)\sqrt{v_1} & \ddots & \ddots & & 0 \\ 0 & \ddots & \ddots & \ddots & & \vdots \\ 0 & 0 & \ddots & \ddots & & \dfrac{n-2n^2}{2} \\ 0 & 0 & 0 & -\sqrt{v_1} & n\left(\dfrac{1}{2}+\sqrt{-E}-\sqrt{v_1}+n\right)+\dfrac{\varepsilon}{4} \end{vmatrix}=0, \qquad (58)$$

which provides important constraints on the potential parameters. Also, from Eq. (3), the wave function is as

$$\psi_n(z)=(1-z)^{\frac{\sqrt{-E_n}}{2}}e^{\frac{\sqrt{v_1}}{2}z}\sum_{m=0}^{n}a_m z^m, \qquad (59)$$

where the expansion coefficients $a_m$ obey the following three-term recurrence relation

$$\left(m^2+\dfrac{3}{2}m+\dfrac{1}{2}\right)p_{m+1}+\left(2\sqrt{v_1}\right)p_{m-1}-\left(m(\chi+m-1)+\dfrac{\varepsilon}{4}\right)p_m=0, \qquad (60)$$

with boundary conditions $a_{-1}=0$ and $a_{n+1}=0$. Now, for comparison with the results of BAM obtained in the previous section, we study the first three states. For $n=0$, from Eq. (57), the energy equation is as

$$(3+2\sqrt{-E_0})\sqrt{v_1}+v_1+v_2=0, \qquad (61)$$

where the constraints on the potential parameters is obtained from (58) as

$$-v_1-v_2-v_3-E_0+\sqrt{-E_0}-\sqrt{v_1}=0. \qquad (62)$$

For the first excited state $n=1$, the energy equation is given by

$$(7+2\sqrt{-E_1})\sqrt{v_1}+v_1+v_2=0, \qquad (63)$$

where the potential parameters satisfy the constraint

$$\left(-v_1-v_2-v_3-E_1+\sqrt{-E_1}-\sqrt{v_1}\right)^2+2\left(-v_1-v_2-v_3-E_1+\sqrt{-E_1}-\sqrt{v_1}\right)\left(3+2\sqrt{-E_1}-2\sqrt{v_1}\right)-8\sqrt{v_1}=0. \qquad (64)$$

Solutions of the second excited state corresponds to $n=2$ are given as

$$(11+2\sqrt{-E_2})\sqrt{v_1}+v_1+v_2=0, \qquad (65)$$

where the constraints on the potential parameters is obtained from (58) as follows

$$\left(-v_1-v_2-v_3-E_2+\sqrt{-E_2}-\sqrt{v_1}\right)^3$$
$$+8\left(-v_1-v_2-v_3-E_2+\sqrt{-E_2}-\sqrt{v_1}\right)^2\left(\frac{13}{4}+\frac{3}{2}\sqrt{-E_2}-\frac{3}{2}\sqrt{v_1}\right)$$
$$+16\left(-v_1-v_2-v_3-E_2+\sqrt{-E_2}-\sqrt{v_1}\right)\left(\frac{\left(3+2\sqrt{-E_2}-2\sqrt{v_1}\right)^2}{2}+3+2\sqrt{-E_2}-6\sqrt{v_1}\right) \quad (66)$$
$$+64\sqrt{v_1}\left(-3-2\sqrt{-E_2}+2\sqrt{v_1}\right)-2\sqrt{v_1}=0.$$

Some numerical results are reported and compared with BAM results in table 1. As can be seen the results achieved by the two methods are identical.

### 3.2. The Razavy bistable potential

In this case, using the results of Appendix, the operator $H$ of Eq. (26) is expressed as an element of the universal enveloping algebra of $sl(2)$ as

$$H\phi(z)=0,$$
$$H=-J_M^+J_M^- + \frac{\xi}{2}J_M^+ + J_M^0 + \frac{\xi}{2}J_M^- + \left(\frac{M-1}{2} + \frac{E+1-2M-\xi^2}{4}\right), \quad (67)$$

where $M=n+1=1,2,3,\ldots$. As a result of the above algebraization, we can use the representation theory of $sl(2)$ which results in a general matrix equation for arbitrary $M$ whose non-trivial solution condition gives the exact solutions of the system as

$$\begin{vmatrix} \frac{E+1-2M-\xi^2}{4} & \frac{\xi}{2} & 0 & 0 & 0 & & 0 \\ \frac{(M-1)\xi}{2} & \frac{E+9-6M-\xi^2}{4} & \xi & 0 & 0 & & 0 \\ 0 & \frac{(M-2)\xi}{2} & \ddots & \ddots & 0 & & 0 \\ 0 & 0 & \ddots & \ddots & \ddots & & 0 \\ 0 & 0 & 0 & \ddots & \ddots & & \frac{(M-1)\xi}{2} \\ 0 & 0 & 0 & 0 & \frac{\xi}{2} & & \frac{E+1-2M-\xi^2}{4} \end{vmatrix}=0. \quad (68)$$

Also, from Eq. (25), the wave function of the system is as

$$\psi_M(z)=z^{\frac{1-M}{2}}e^{-\frac{\xi}{4}\left(z+\frac{1}{z}\right)}\sum_{m=0}^{M-1}a_m z^m, \quad (69)$$

where the coefficients $a_m$ satisfy the following recursion relation

$$\left(\frac{m\xi}{2}\right)a_m + \left(\frac{E-\xi^2-6m+3}{4}\right)a_{m-1} + (\xi)a_{m-2}=0, \quad (70)$$

with initial conditions $a_M = 0$ and $a_{-1} = 0$. As examples of the general formula (68) and also, for comparison purpose, we study the first three levels. For the ground state $M = 1$, from (68), we have

$$E_0 = 1 + \xi^2. \tag{71}$$

For the first excited state, from Eq. (68), we obtain

$$E_2 = \xi^2 + 3 \pm 2\xi. \tag{72}$$

In a similar way, for the second excited state $M = 3$, we have

$$\frac{1}{64}\left(\xi^4 + (-2E_3 - 2)\xi^2 + E_3^2 - 14E_3 + 45\right)\left(-\xi^2 + E_3 - 5\right) = 0, \tag{73}$$

which yields the energy as

$$E_3 = \begin{cases} \xi^2 + 5 \\ \xi^2 + 7 \pm 2\sqrt{4\xi^2 + 1} \end{cases}. \tag{74}$$

The numerical results are reported and compared with BAM results in table 2.

### 3.3. The hyperbolic Shifman potential

Comparing Eq. (42) with the results of Appendix, it is seen that the operator $H$ can be expressed in the Lie-algebraic form

$$H\phi(z) = 0,$$
$$H = \frac{1}{2}\left(J_n^+ J_n^- + J_n^- J_n^-\right) - a\left(J_n^+ + J_n^-\right) - \left(\frac{n+1}{2}\right)J_n^0 - \left(E + \frac{n(n+1)}{4}\right). \tag{75}$$

Then, using the representation theory of $sl(2)$, the general condition for existence of a non-trivial solution is obtained as

$$\begin{vmatrix} -E & -a & 1 & 0 & 0 & 0 \\ -na & -E - \frac{1}{2} & -2a & 3 & 0 & \vdots \\ 0 & -(n-1)a & \ddots & \ddots & \ddots & 0 \\ 0 & 0 & \ddots & \ddots & \ddots & \frac{n(n-1)}{2} \\ \vdots & 0 & 0 & \ddots & \ddots & -na \\ 0 & \cdots & \cdots & 0 & -a & -E - \frac{n^2}{2} \end{vmatrix} = 0, \tag{76}$$

which gives the same results as those obtained by BAM in the previous section. For example, for $n = 0$, from Eq. (76), the ground state energy of the system is $E_0 = 0$. For $n = 1$, the energy equation is as follows

$$2E_1^2 + E_1 - 2a^2 = 0, \tag{77}$$

which yields

$$E_1 = \frac{-1 \pm \sqrt{16a^2 + 1}}{4}. \tag{78}$$

Likewise, for $n = 2$, the second excited state energy of the system is obtained from the following relation

$$-E_2^3 - \frac{5}{2}E_2^2 + (4a^2 - 1)E_2 + 6a^2 = 0. \tag{79}$$

The numerical results are reported and compared with BAM results in table 4. The wave function of the system from Eq. (41), is given as

$$\psi_n(z) = e^{-az} \sum_{m=0}^{n} a_m z^m, \tag{80}$$

where the expansion coefficients $a_m$'s satisfy the recursion relation

$$(-(m+1)a)a_{m-2} - \left(E + \frac{m-1}{2}\right)a_{m-1} - (ma)a_m + \left(\frac{m(m+1)}{2}\right)a_{m+1} = 0, \tag{81}$$

with boundary conditions $a_{-2} = 0$, $a_{-1} = 0$ and $a_{n+1} = 0$.

**Conclusions**

Using the Bethe ansatz method, we have solved the Schrödinger equation for a class of QES DWPs and obtained the general expressions for the energies and the wave functions in terms of the roots of the Bethe ansatz equations. In addition, we have solved the same problems using the Lie algebraic approach within the framework of quasi-exact solvability and obtained the exact solutions using the representation theory of $sl(2)$ Lie algebra. It was found that the results of the two methods are consistent. Also, we have provided some numerical results for the Razavy potential to display the energy splitting explicitly. The main advantage of the methods we have used is that we determine the general exact solutions of the systems and thus, we can quickly calculate the solution of any arbitrary state, without cumbersome procedures and without difficulties in obtaining the solutions of the higher states.

**Appendix A: The Lie algebraic approach of quasi-exact solvability**

In this Appendix, we outline the connection between $sl(2)$ Lie algebra and the second order QES differential equations. A differential equation is said to be QES if it lies in the universal enveloping algebra of a QES Lie algebra of differential operators [12]. In the case of one-dimensional systems, $sl(2)$ Lie algebra is the only algebra of first-order differential operators that possesses finite-dimensional representations, whose generators [12]

$$J_n^+ = -z^2 \frac{d}{dz} + nz,$$

$$J_n^0 = z\frac{d}{dz} - \frac{n}{2}, \qquad (A\text{-}1)$$

$$J_n^- = \frac{d}{dz},$$

obey the $sl(2)$ commutation relations as

$$[J_n^+, J_n^-] = 2J_n^0, \quad [J_n^0, J_n^\pm] = \pm J_n^\pm, \qquad (A\text{-}2)$$

and leave invariant the finite-dimensional space

$$P_{n+1} = \langle 1, z, z^2, ..., z^n \rangle. \qquad (A\text{-}3)$$

Hence, the most general one-dimensional second order differential equation $H$ can be expressed as a quadratic combination of the $sl(2)$ generators as

$$H = \sum_{a,b=0,\pm} C_{ab} J^a J^b + \sum_{a=0,\pm} C_a J^a + C. \qquad (A\text{-}4)$$

On the other hand, the operator (A-4) as an ordinary differential equation has the following differential form

$$H\phi(z) = 0,$$

$$H = -P_4(z)\frac{d^2}{dz^2} + P_3(z)\frac{d}{dz} + P_2(z), \qquad (A\text{-}5)$$

where $P_l$ are polynomials of at most degree $l$. Generally, this operator does not have the form of Schrödinger operator but can always be turned into a Schrödinger-like operator

$$\tilde{H} = e^{-A(x)} H e^{A(x)} = -\frac{1}{2}\frac{d^2}{dx^2} + (A')^2 - A'' + P_2(z(x)), \qquad (A\text{-}6)$$

using the following transformations

$$x = \pm \int \frac{dz}{\sqrt{P_4}},$$

$$\phi(z) = e^{-\int \left(\frac{P_3}{P_4}\right) dz + \log x'} \psi(x). \qquad (A\text{-}7)$$

The interested reader is referred to Refs. [9-14] for further details.

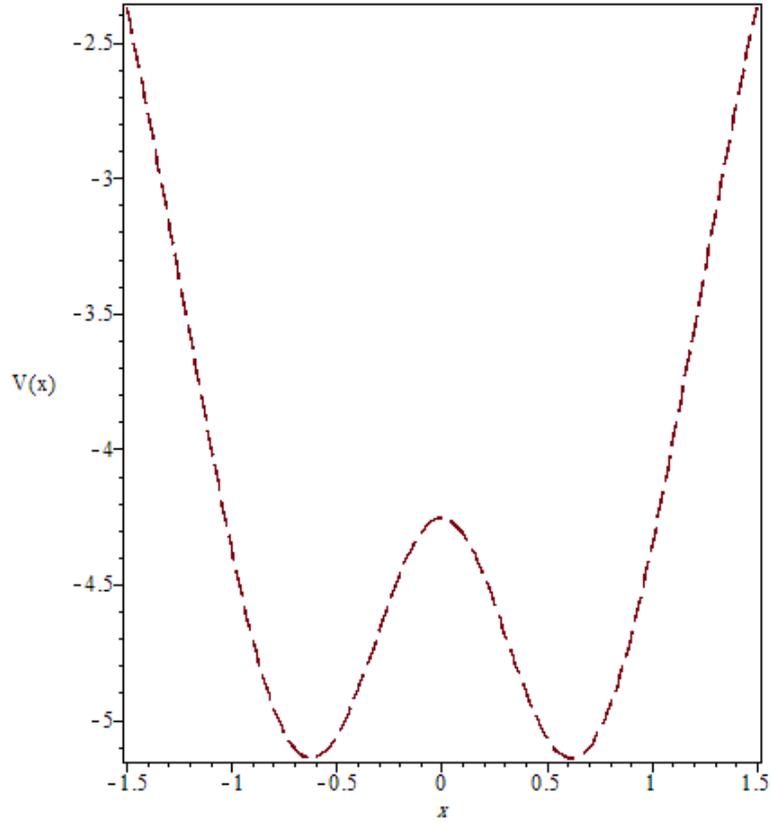

**Fig.1.** The generalized Manning DW potential with $v_1 = 1$, $v_2 = -12$ and $v_3 = 15.255$.

| $n$ | $v_2$ | Energy (BAM) Eq. (12) | $v_3$ (BAM) Eq.(13) | Energy (QES) Eq. (57) | $v_3$ (QES) Eq.(58) | Energy Ref. [36] | Wave function (BAM) |
|---|---|---|---|---|---|---|---|
| 0 | $-6$ | $-1$ | 6 | $-1$ | 6 | $-1$ | $\psi_0(z) = (1-z)^{\frac{\sqrt{-E_0}}{2}} e^{\frac{1}{2}z}$ |
| 1 | $-12$ | $-4$ | 15.2554<br>26.7446 | $-4$ | 15.2554<br>26.7446 | $-$ | $\psi_1(z) = \begin{cases} (1-z)^{\frac{\sqrt{-E_1}}{2}} e^{\frac{1}{2}z} (z+2.6862) \\ (1-z)^{\frac{\sqrt{-E_1}}{2}} e^{\frac{1}{2}z} (z-0.1862) \end{cases}$ |
| 2 | $-18$ | $-9$ | 26.8458<br>41.1214<br>66.0329 | $-9$ | 26.8458<br>41.1214<br>66.0329 | $-$ | $\psi_2(z) = \begin{cases} (1-z)^{\frac{\sqrt{-E_2}}{2}} e^{\frac{1}{2}z} (z^2 + (9.2886)z + 16.0951) \\ (1-z)^{\frac{\sqrt{-E_2}}{2}} e^{\frac{1}{2}z} (z^2 + (5.7197)z - 0.8718) \\ (1-z)^{\frac{\sqrt{-E_2}}{2}} e^{\frac{1}{2}z} (z^2 - (0.5082)z + 0.0267) \end{cases}$ |

**Table 1.** Solutions of the first three states for the generalized Manning DWP with $v_1 = 1$ and the possible values of $v_2$, where $z = \tanh^2(x)$.

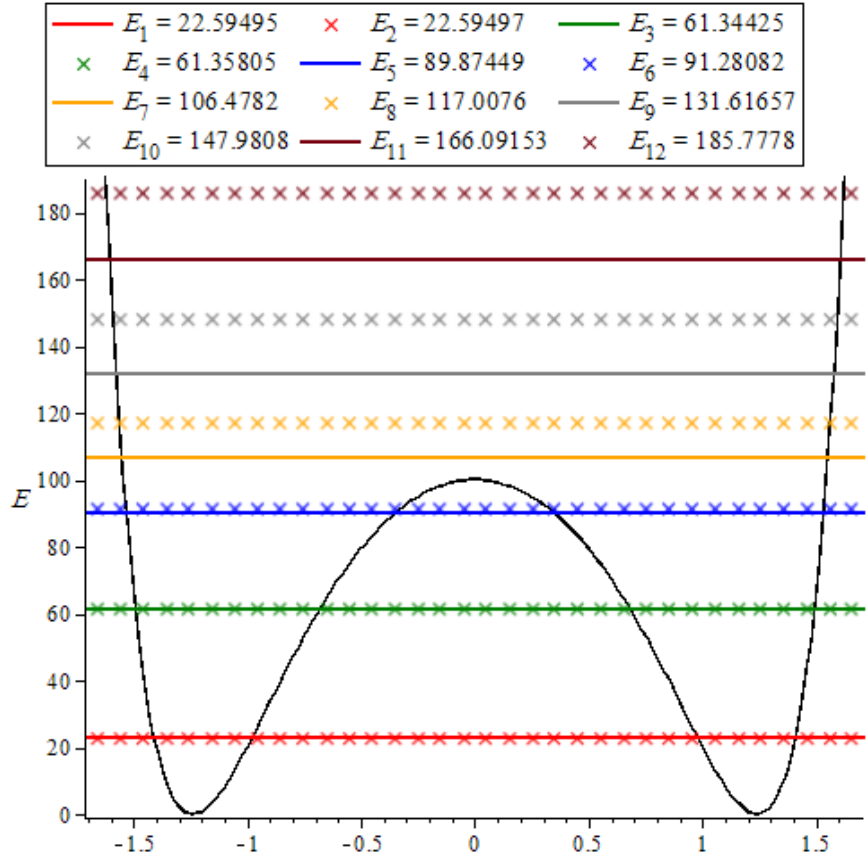

**Fig.2.** The Razavy bistable potential and the energy levels splitting with $\xi = 2$ and $M = 12$.

| M | Energy (BAM) Eq. (30) | Energy (QES) Eq. (68) | Energy Ref. [44] | Wave function (BAM) |
|---|---|---|---|---|
| 1 | 5 | 5 | 5 | $\psi_1(z) = e^{-\frac{1}{2}\left(z+\frac{1}{z}\right)}$ |
| 2 | 3<br>11 | 3<br>11 | 3<br>11 | $\psi_2(z) = \begin{cases} z^{\frac{-1}{2}} e^{-\frac{1}{2}\left(z+\frac{1}{z}\right)}(z+1) \\ z^{\frac{-1}{2}} e^{-\frac{1}{2}\left(z+\frac{1}{z}\right)}(z-1) \end{cases}$ |
| 3 | 2.75389<br>19.2462 | 2.75389<br>19.2462 | - | $\psi_3(z) = \begin{cases} z^{-1} e^{-\frac{1}{2}\left(z+\frac{1}{z}\right)}\left(z^2 + (1.5616)z + 1\right) \\ z^{-1} e^{-\frac{1}{2}\left(z+\frac{1}{z}\right)}\left(z^2 - 1\right) \\ z^{-1} e^{-\frac{1}{2}\left(z+\frac{1}{z}\right)}\left(z^2 - (2.5616)z + 0.9999\right) \end{cases}$ |

**Table 2.** Solutions of the first three states for the Razavy bistable potential with $\xi = 2$, where $z = e^{2x}$.

| $M$ | $E$ | $\Delta E$ |
|---|---|---|
| 1 | 5 | |
| 2 | 3<br>11 | $E_2 - E_1 = 8$ |
| 3 | 2.753788749<br>9<br>19.24621125 | $E_2 - E_1 = 6.246211251$ |
| 4 | 4.071796770<br>8.416994756<br>17.92820323<br>29.58300524 | $E_2 - E_1 = 4.345197986$<br>$E_4 - E_3 = 11.65480201$ |
| 5 | 6.541491983<br>9<br>18.46929468<br>29<br>41.98921334 | $E_2 - E_1 = 2.458508017$<br>$E_4 - E_3 = 10.53070532$ |
| 6 | 9.410470052<br>10.41150119<br>21.29169652<br>30.14478972<br>42.29783343<br>56.44370909 | $E_2 - E_1 = 1.001031138$<br>$E_4 - E_3 = 8.85309320$<br>$E_6 - E_5 = 14.14587566$ |
| 7 | 12.00197331<br>12.28436662<br>26.60488323<br>33<br>44.45652679<br>57.71563338<br>72.93661667 | $E_2 - E_1 = 0.28239331$<br>$E_4 - E_3 = 6.39511677$<br>$E_6 - E_5 = 13.25910659$ |
| 8 | 14.2739943644243<br>14.3338521331318<br>33.7279836821029<br>37.3692217631865<br>48.7528515817037<br>60.8366196369519<br>75.2451703717691<br>91.4603064667298 | $E_2 - E_1 = 0.05985777$<br>$E_4 - E_3 = 3.64123808$<br>$E_6 - E_5 = 12.08376806$<br>$E_8 - E_7 = 16.21513610$ |
| 9 | 16.4065712724492<br>16.4168634672976<br>41.3677353353592<br>42.8331930589016<br>55.7615205763811<br>65.8916709587647<br>79.4544486908676<br>94.8582725150361<br>112.009724124943 | $E_2 - E_1 = 0.01029220$<br>$E_4 - E_3 = 1.46545772$<br>$E_6 - E_5 = 10.13015038$<br>$E_8 - E_7 = 15.40382383$ |
| $M$ | $E$ | $\Delta E$ |
| 10 | 18.4886350611369<br>18.4901336128637<br>48.4784563542366<br>48.8807648437805<br>65.7434045220649<br>72.8690185914347<br>85.7451955766933<br>100.179116457372<br>116.544308485868<br>134.580966494549 | $E_2 - E_1 = 0.00149855$<br>$E_4 - E_3 = 0.40230849$<br>$E_6 - E_5 = 7.12561407$<br>$E_8 - E_7 = 14.43392092$<br>$E_{10} - E_9 = 18.0366580$ |
| 11 | 20.54831894<br>20.54850863<br>55.03511334<br>55.11772987<br>77.68896865<br>81.50920136<br>94.52213923<br>107.5333540<br>123.0343674<br>140.2912061<br>159.1710925 | $E_2 - E_1 = 0.00018969$<br>$E_4 - E_3 = 0.08261653$<br>$E_6 - E_5 = 3.82023271$<br>$E_8 - E_7 = 13.01121477$<br>$E_{10} - E_9 = 17.2568387$ |
| 12 | 22.59494691<br>22.59496818<br>61.34425227<br>61.35805469<br>89.87448537<br>91.28081517<br>106.4782162<br>117.0076415<br>131.6165721<br>147.9807662<br>166.0915272<br>185.7777543 | $E_2 - E_1 = 0.00002127$<br>$E_4 - E_3 = 0.01380242$<br>$E_6 - E_5 = 1.40632980$<br>$E_8 - E_7 = 10.5294253$<br>$E_{10} - E_9 = 16.3641941$<br>$E_{12} - E_{11} = 19.6862271$ |

**Table 3.** Energy splitting of the first 12 states for the Razavy bistable potential with $\xi = 2$.

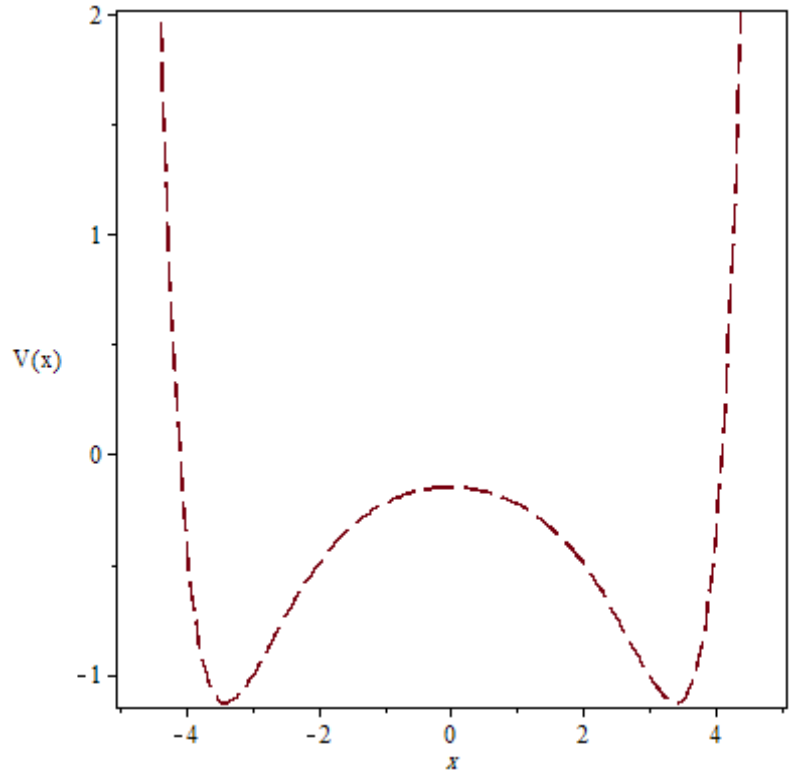

**Fig. 3.** The Shifman DWP with $a = 0.1$ and $n = 1$

| $n$ | Energy (BAM) Eq. (46) | Energy (QES) Eq. (76) | Energy Ref. [9] | Wave function (BAM) |
|---|---|---|---|---|
| 0 | 0 | 0 | 0 | $\psi_0(z) = e^{-\frac{1}{10}z}$ |
| 1 | 0.0193<br>−0.5193 | 0.0193<br>−0.5193 | 0.0193<br>−0.5193 | $\psi_1(z) = \begin{cases} e^{-\frac{1}{10}z}(z - 5.1926) \\ e^{-\frac{1}{10}z}(z + 0.1926) \end{cases}$ |
| 2 | −2.0067<br>0.05457<br>−0.5479 | −2.0067<br>0.05457<br>−0.5479 | −2.0067<br>0.05457<br>−0.5479 | $\psi_2(z) = \begin{cases} e^{-\frac{1}{10}z}(z^2 + (0.0670)z - 0.4949) \\ e^{-\frac{1}{10}z}(z^2 - (20.5457)z + 55.9707) \\ e^{-\frac{1}{10}z}(z^2 - (14.5213)z - 4.4757) \end{cases}$ |

**Table 4.** Solutions of the first three states for the Shifman DWP with $a = 0.1$, where $z = \cosh(x)$.